**High-efficiency silicon LED with ultra-wideband emission from visible to infrared at room temperature**

Xiaobo Li[1,2,#], Jiajing He[1,2, #*], Yaping Dan[3], Jun Wang[1,2]

# These authors contributed equally to this work

[1]Aerospace Laser Technology and System Department, Shanghai Institute of Optics and Fine Mechanics, Chinese Academy of Sciences, Shanghai 201800, China.

[2]Center of Materials Science and Optoelectronics Engineering, University of Chinese Academy of Sciences, Beijing 100049, China

[3]University of Michigan – Shanghai Jiao Tong University Joint Institute, Shanghai Jiao Tong University, Shanghai 200240, China

**Corresponding author**: Jiajing He, Shanghai Institute of Optics and Fine Mechanics, Chinese Academy of Sciences, No 390, Qinghe Road, 201800, Shanghai, China; University of Chinese Academy of Sciences, Beijing 100049, China

**E-mail**: jiajinghe@siom.ac.cn

**Abstract**

The primary challenge in silicon photonics is achieving efficient luminescence in the communication band, crucial for its large-scale application. Despite significant efforts, silicon light sources still suffer from low efficiency and limited emission wavelengths. We addressed this by achieving broadband luminescence from 600-1650 nm through femtosecond laser annealing of 220nm standard SOI, resulting in an external quantum efficiency exceeding 0.26% and an output optical power density greater than 20 W/cm², several orders of magnitude higher than other silicon-based LEDs in performance. The broadband LED has potential applications in optical inspection, gas sensing, optical coherence tomography, optical communication, and more.

## 1. INTRODUCTION

Silicon photonics enable the miniaturization and mass production of optical systems to meet the increasing demand across various applications. At present, silicon photonics systems require the integration of external light sources, which will increase the complexity and cost of the system[1]. As technology develops towards high integration and miniaturization, the demand for on-chip light sources becomes more urgent, and the limitations of silicon in this regard need to be urgently addressed. However, as an indirect bandgap semiconductor, the light emission process of silicon involves a complex non-radiative transition. Electrons interact with lattice vibrations to satisfy momentum conservation requirements besides transition between energy states, which will greatly increases greatly increases energy loss, thus seriously restricting the light-emitting efficiency of silicon[2]. Although various optoelectronic devices have been successfully integrated into microelectronic platforms, promoting the development of optoelectronic technology, the development of on-chip light sources still faces many challenges. In the search for solutions, incremental advances have been made through techniques such as wafer bonding and III-V semiconductor growth, which offer high output power and meet communication wavelengths, thereby revitalizing hope for silicon photonics[3,4]. However, these technologies are still difficult to integrate into the standard CMOS platform and can substantially raise processing costs. Although CMOS-compatible silicon light-emitting diode (LED) technology has been widely explored, such as electroluminescence in forward-biased diodes and luminescence based on collision ionization in reverse-biased P-(I)-N diodes, these technologies have problems such as low electro-optical conversion efficiency, low light power density, and incompatibility with communication wavelengths[5-7].

Silicon photonics technology has significant application potential beyond communications, including quantum computing and sensing. While some refractive index sensing relies on fixed-wavelength lasers, the advancement of techniques such as Fourier transform infrared spectroscopy, waveguide infrared absorption spectroscopy, and micro-ring absorption spectroscopy requires low-cost ultra-broadband light sources[8-11]. Previously, supercontinuum light sources with ultra-high working bandwidth have been experimentally verified in silicon nitride and silicon, bringing new hope for the development of silicon photonics[12,13]. However, the practical application of this technology still faces many difficulties. On the one hand, the generation of supercontinuum light sources often relies on high-power continuous wave pump lasers or femtosecond lasers, which are large, costly, and energy-intensive, hindering the development of compact, low-power systems[14,15]. On the other hand, the lack of supercontinuum light sources that can be integrated on a single chip complicates achieving high integration and portability across optoelectronic systems. These limitations severely restrict silicon photonics applications in areas such as wearable devices and mobile terminals, which demand strict size and power consumption requirements. Consequently, in-depth research into the efficient light-emission mechanisms of silicon and exploring their potential in light emission are essential for advancing semiconductor technology.

To address these issues, we use femtosecond laser annealing (FLA) technology on a 220 nm silicon-on-insulator (SOI) substrate. By precisely controlling the process parameters of FLA, we successfully achieved broadband photoluminescence and electroluminescence in the range of 600 to 1650 nm, with a photoluminescent external quantum efficiency (EQE) close to 0.3% (in the range of 900 nm to 1650 nm). The output power of the LED exceeds 26 μW, and the maximum EQE exceeding 0.26% (900 nm to 1650 nm). Additionally, the output optical power density reaches 20

W/cm$^2$, significantly outperforming other silicon-based LEDs by several orders of magnitude. Moreover, this technique is fully compatible with CMOS technology, significantly enhancing the operating bandwidth while maintaining high light emission efficiency. As a result, it can be widely used in data communication, chip defect detection, wearable devices, sensing[16], as well as infrared imaging for night vision and hyperspectral imaging[17].

## 2. RESULTS

We measure the power-dependent photoluminescence (PL) spectra of silicon after FLA at room temperature (Fig. 1a) and observe ultra-wideband luminescence in the range of 900-1650 nm. The fluorescence intensity significantly increases with rising excitation power. The attenuation of the PL signal below 1000 nm and above 1600 nm is due to the limitations of the InGaAs detector, the actual emission range is broader, as shown in Fig.S1. Fig. 1b presents the slope efficiency of the infrared emission integral intensity as a function of excitation power (on a logarithmic-logarithmic scale), with values of 5.40 and 9.62, both significantly greater than 1 and exceeding those reported for materials such as silicon[18,19], InP[20] and SiGe[21]. This indicates that, even at room temperature, the sample exhibits ultra-high internal quantum efficiency, with minimal effects from non-radiative recombination processes such as Auger or heat-dependent processes[22]. The inset in Fig. 1b illustrates the nonlinear relationship between output power and pump power, indicating that within a specific power range, increasing the pump power leads to a significant rise in output power. Fig. 1c presents the relationship between the EQE of silicon treated with FLA and excitation power in the range of 900 to 1650 nm, with data derived from calibration using Hamamatsu commercial LED L12509-0155G [23]. As the excitation power increases, the EQE rises significantly, with its slope gradually increasing, reaching a maximum EQE of nearly 0.3%, which is four orders of magnitude higher than that of pure silicon ($10^{-7}$)[18], and close to the direct band gap SiGe alloys[21]. It is important to note that the current EQE is far from saturation (the slope has not diminished), indicating that there is potential for further increases in EQE with additional increases in excitation power. Furthermore, we have only calculated the EQE in the 900 to 1650 nm range, considering its strong emission capability in the visible range, the EQE over the entire spectrum will be much greater than 0.3% (see Fig.S1). Fig. 1d displays the fluorescence lifetime of the samples before and after FLA at room temperature. It is clear that the silicon after FLA has a longer fluorescence lifetime.

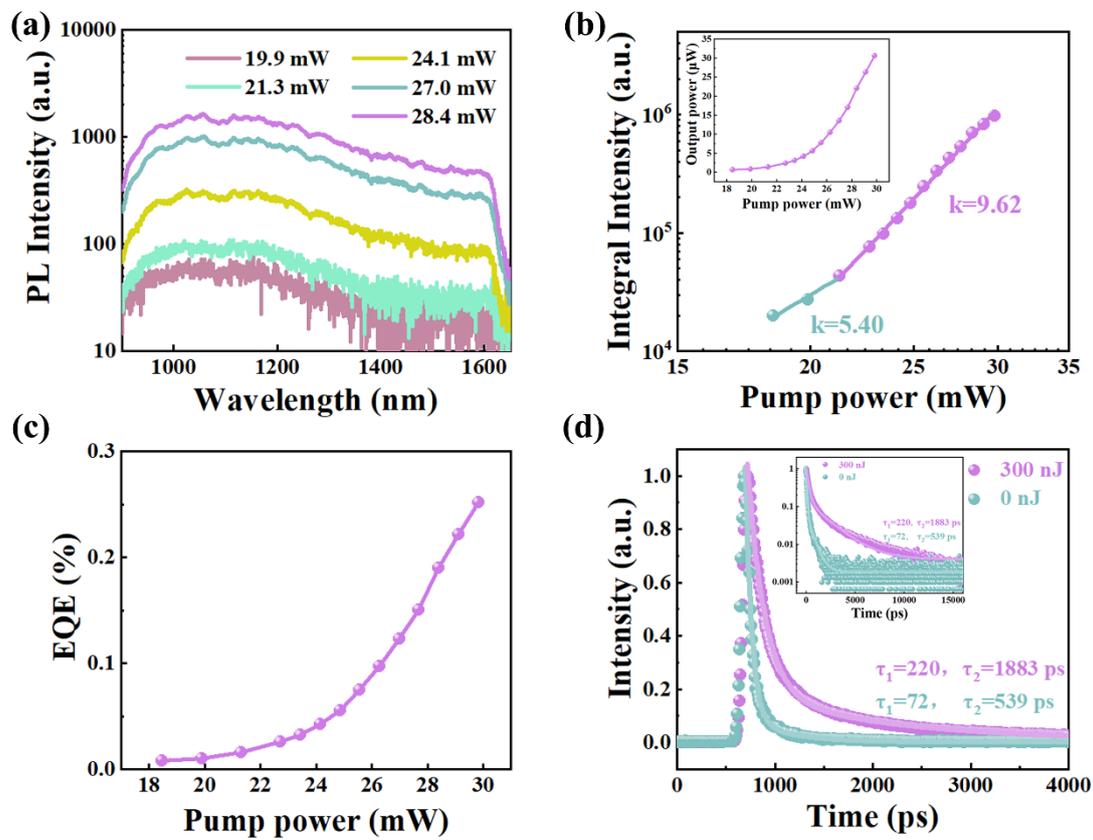

**Fig. 1 | PL test of the sample. a,** PL spectra of infrared light excited by varying powers. **b,** Relationship between PL integral intensity of infrared light and pump power, with the straight line representing the linear fit of PL integral intensity with variable pump power (log-log scale), along with an illustration of the output power versus injection current. The illustration of the correlation between output power and pump power is included. **c,** EQE of infrared light at varying pump power. **d,** Fluorescence lifetime of silicon after FLA (532nm excitation).

The FLA silicon sample demonstrated significant broadband luminescence in the infrared region. To explore the underlying mechanisms, we analyze its infrared carrier dynamics using femtosecond transient absorption (TA) measurement, as shown in Fig. 2. Fig. 2a illustrates the TA spectroscopy using a pump wavelength of 515 nm and a detection range of 1200 nm to 1700 nm. The spectrum reveals a relatively broad absorption profile with several distinct peaks, indicating that the absorption wavelength of silicon extends into the infrared region after FLA[24]. Additionally, we observe the induced bleaching signal peak in the range of 1500 nm to 1550 nm (dashed line), possibly due to the Pauli blocking of excited electrons. The decay-related TA spectra reveal more detailed carrier dynamics (Fig. 2b) with time constants of 0.5 ps, 1.1 ps, 2.2 ps, 3.2 ps, 10.3 ps, and 97.6 ps. Significant absorption signal changes occur at short delays (0.5-3.2 ps), indicating rapid excited-state evolution; at longer delays (10.3-97.6 ps), changes slow down, indicating the excited state stabilizes.

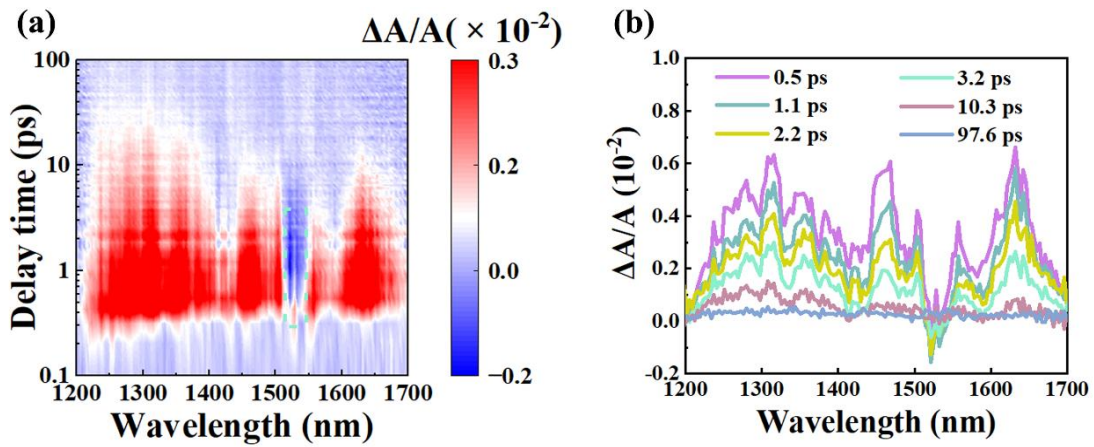

**Fig. 2 | TA measurement of silicon after FLA. a,** Infrared TA diagram of silicon after FLA. **b,** TA spectra at different decay times.

Leveraging the excellent PL characteristics of silicon after FLA, we have designed a broadband LED, as shown in Fig. 3a. The manufacturing process is fully compatible with CMOS technology and involves multiple lithography and ion implantation steps to create horizontal PIN junctions, which are then annealed with FLA to produce an electroluminescent wideband LED (details in the Device fabrication section). Unlike other broadband LED that operates through reverse avalanche breakdown[25], our LED achieves electroluminescence by applying a forward bias to the horizontal PIN junction. Fig. 3b shows the current-correlation EL spectrum of the PIN junction at room temperature, indicating ultra-broadband emission similar to PL in the 900 to 1650 nm range. Additionally, Fig. S2 shows that the EL also has strong emission in the visible range, thus generating broadband emission from 700 to 1650 nm in our LED. Fig. 3c illustrates the relationship between the integrated intensity of the LED and the driving current on a log-log scale. The slope is 3.64 at low currents but increases to 6.08 after reaching the threshold current of 9 mA. Notably, the slopes in this range exceed 1, indicating the LED's potential as a laser. We subsequently present the EQE of infrared emission at different injection currents, as shown in Fig. 3d. The EQE exhibits super-linear behavior as the current increases, with higher currents within a specific range resulting in greater EQE. At a bias current of 11.4 mA, the output power reaches 28 μW, the output power density exceeded 20 W/cm$^2$, and the EQE surpasses 0.26%. Notably, this EQE is calculated for emissions in the 900-1650 nm range, considering a broader emission range (700-1650nm), the total EQE is significantly higher. It is truly remarkable that silicon, an indirect bandgap semiconductor, can produce such broad and efficient light emission.

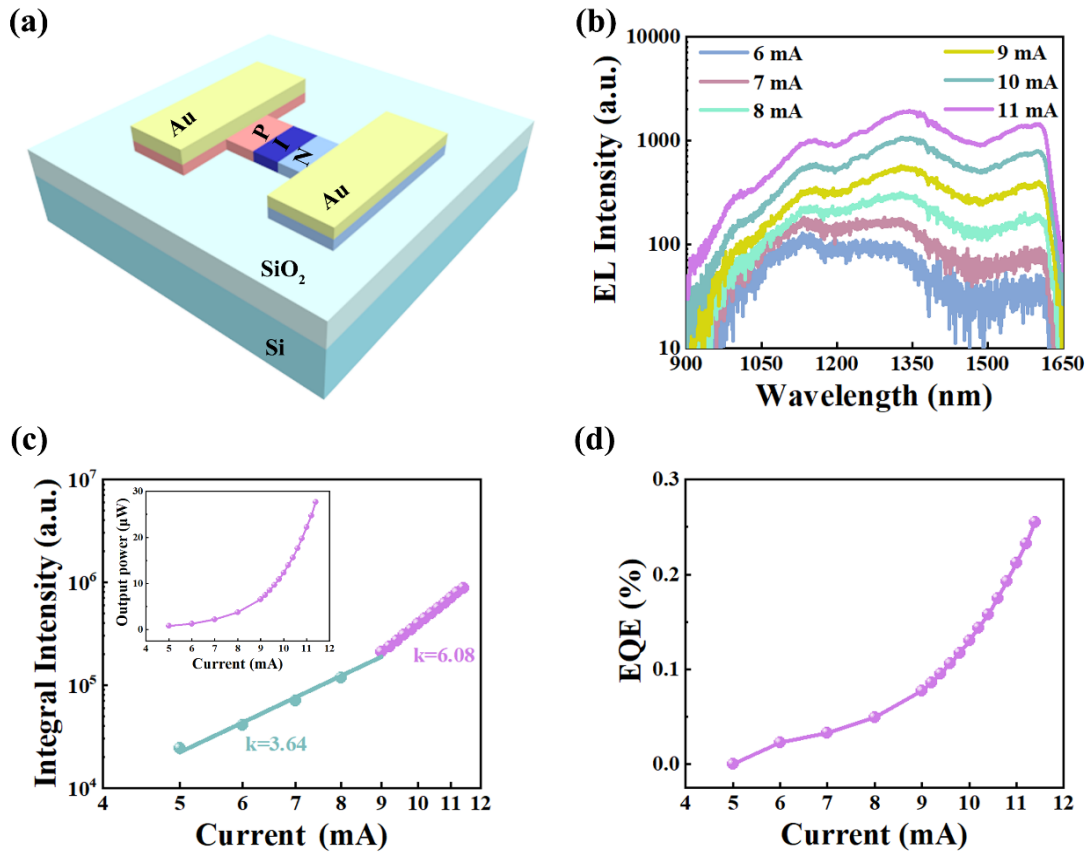

**Fig. 3 | Electroluminescence of the sample. a,** Schematic diagram of a broadband silicon LED. **b,** EL spectrum of infrared light at different injection currents. **c,** Relationship between infrared integral EL intensity and injection current, along with an illustration of the output power versus injection current. **d,** Relationship between EQE and injection current.

Compared to previously reported silicon-based LEDs (Fig. 4) [25-29], our LED exhibits a superior combination of large output bandwidth, high EQE and optical power density. Table 1 compares the luminescent performance of our silicon-based LED with other LEDs. Our LED achieves several orders of magnitude improvements in output power density and EQE, significantly surpassing the band edge luminescence of silicon.

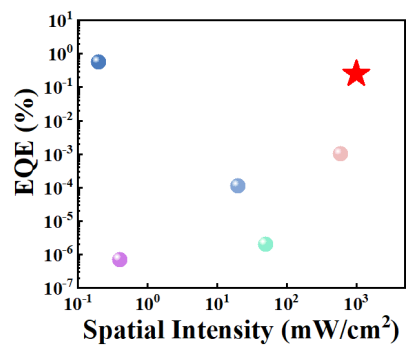

**Fig. 4 | Comparison of the LED's EQE and output optical power density with related work.**

**Table 1** Comparison with other silicon-based LEDs. n/a: not available.

| Reference | Wavelength (nm) | Power (μW) | Power Density (mW/cm$^2$) | EQE |
|---|---|---|---|---|
| This Work | 600-2200 | >28 (11.4mA) | 20000 | >0.26% |
| Pruessner (2024)[25] | 900-1600 | 5.1× 10$^{-3}$ (5mA) | 4100 (post-focusing) | 1.1× 10$^{-6}$ |
| Wang (2024)[26] | 1500-1600 | <0.1 (10mA) | n/a | 8× 10$^{-6}$ |
| Green (2001)[27] | 1050-1250 | 180 (40mA) | <0.2 | 0.55% |
| Ng (2001)[28] | 1050-1250 | 19.8 (100mA) | n/a | 2× 10$^{-4}$ |
| Li (2023)[29] | 1000-1200 | 9.4× 10$^{-5}$ (6mA) | >50 | <2× 10$^{-8}$ |

## 3. CONCLUSIONS

We achieve ultra-broadband PL on silicon from 600 to 1650 nm (532 nm excitation) with an EQE close to 0.3%. TA reveals the carrier relaxation process, confirming the absorption extension. Subsequently, we fabricate a 900-1650 nm ultra-broadband LED on 220 nm standard SOI, fully compatible with CMOS processes, with emission wavelengths compatible with communication requirements. Under a forward bias of 11.4 mA, the output power reaches 28 μW, the optical power density exceeds 20 W/cm², and the EQE surpasses 0.26%. Its performance is several orders of magnitude higher than other silicon-based LEDs, making it highly valuable for applications in optical detection, wearable devices, gas sensing, optical coherence tomography, and optical communication.

## Methods

**FLA**

we use the FemtoLAB system from UAB Altechna R&D for femtosecond laser annealing. The system features a Pharos femtosecond laser with a central wavelength of 515 nm, a pulse width of 290 fs, and repetition rate of 100 kHz, allowing precise control of laser energy through attenuation filters. The laser beam is focused onto the sample surface with a 10× objective lens (NA=0.26), delivering 300 nJ of energy for the annealing process.

**Device fabrication**

We form a PN junction by locally ion-implanting boron (20 keV, $9\times10^{14}$ cm$^{-2}$) and phosphorus (60 keV, $9\times10^{14}$ cm$^{-2}$). After rapid annealing at 950°C for 30 seconds to activate the implanted ions, we perform FLA in the I region.

**Photoluminescence/Electroluminescence Testing**

Fluorescence signals are collected using the WITec confocal Raman system Alpha300R. The system has an excitation wavelength of 532 nm and employs a 10× objective lens (NA=0.25). The visible light diffraction grating is 300 g/mm (BLZ 750.00 nm), while the infrared grating is 150 g/mm (BLZ 1250.00 nm).

**Transient Absorption Spectroscopy Testing**

Microregion testing is conducted using the TA100-1030NM-NIR-MIC ultrafast transient absorption spectroscopy system from the time tech spectra, which features a maximum time window of 8 ns. The primary pump light (515 nm, 100 kHz) excites carriers from the ground state to an excited state. A weaker, time-delay-tunable white light (1200-1700 nm) is then used for detection at the location where the pump light overlaps with the sample surface.

**Extraction of EQE**

We used the WITec Alpha300R system to measure the photoluminescent and electroluminescent EQE of silicon samples treated with FLA. In this process, accurately calibrating the total loss caused by all components in the optical path is crucial for ensuring the reliability of the test results. Typically, we estimate the total loss by assuming the optical path conditions and accumulating the known loss coefficients of each component. However, due to significant differences in optical path design, component parameters, and environmental conditions across different experiments, this method often leads to substantial deviations between the estimated values and the actual situation. To overcome this challenge and simplify the measurement process, we adopte a method based on the intensity ratio of a traditional LED to the sample's photoluminescence (PL) [23,30-32]. Specifically, we select the Hamamatsu LED (L12509-0155G) as the reference standard for the system's total loss. During the measurement, to ensure accuracy and comparability, we incorporate an appropriate infrared attenuator into the Alpha300R system to attenuate the fluorescent signal from the standard Hamamatsu LED to a level comparable to that of the infrared fluorescent signal from the sample under test. Subsequently, by precisely measuring and comparing the PL/EL spectral areas of the standard LED and the sample, we can accurately calculate the optical power of the sample using the mathematical relationships between them. Once we obtain the optical power data, we combine it with relevant parameters such as the pump light power and, based on the definition and calculation formula of EQE, accurately determine the photoluminescent and electroluminescent EQE of the sample under test.

$$EQE = \frac{P_{out}}{P_{in}}$$

**Visible light photoluminescence characteristics**

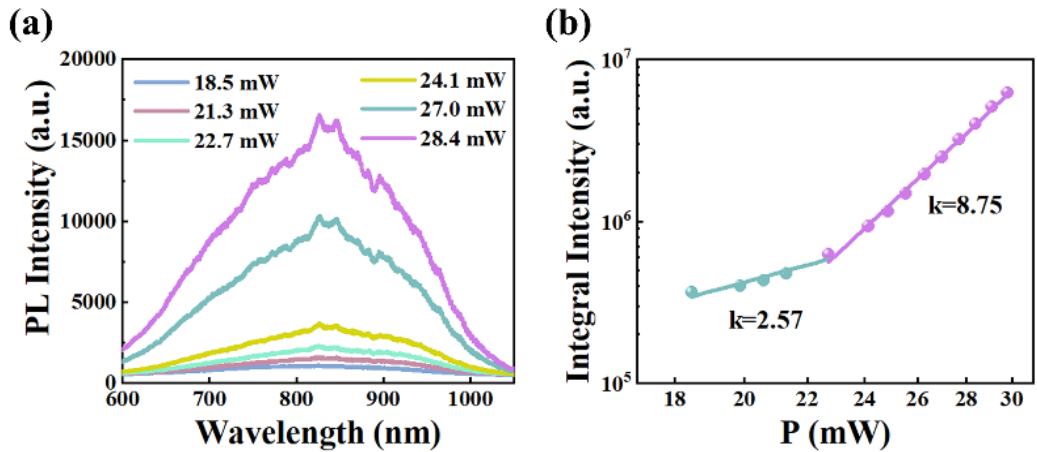

**Fig. S1 a,** Photoluminescence spectra of visible light excited by varying powers. **b,** Relationship between photoluminescence integral intensity of visible light and pump power, with the straight line representing the linear fit of photoluminescence integral intensity with variable pump power (log-log scale)

**Visible light electroluminescence characteristics**

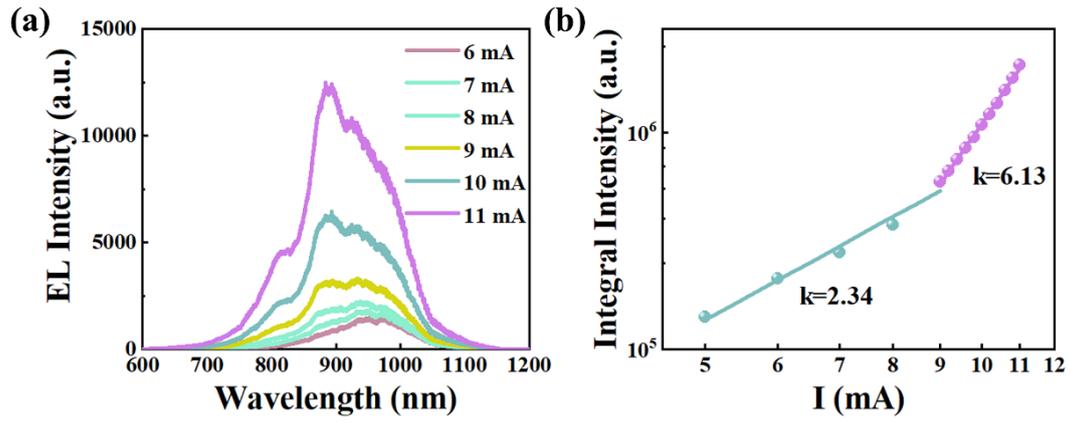

**Fig. S2 Visible light EL characterization of the LED. a,** Visible light EL spectrum under different injection currents. **b,** Relationship between the integral EL intensity of visible light and the injection current (log scale)